\documentclass[prb,letterpaper,aps,superscriptaddress,floatfix,twocolumn]{revtex4-1}
\usepackage{graphicx}
\usepackage{amsmath}
\usepackage{subfigure}
\newcommand{\beq}{\begin{equation}}
\newcommand{\eeq}{\end{equation}}
\newcommand{\bk}{{{\bf{k}}}}

\newcommand{\br}{{{\bf{r}}}}
\newcommand{\bR}{{{\bf{R}}}}

\newcommand{\bA}{{\bf{A}}}
\newcommand{\bB}{{\bf{B}}}

\newcommand{\ba}{{\bf{a}}}

\newcommand{\bq}{{\bf{q}}}
\newcommand{\bp}{{\bf{p}}}

\newcommand{\bn}{{\bf{n}}}
\newcommand{\bbm}{{\bf{m}}}

\newcommand{\beqa}{\begin{eqnarray}}
\newcommand{\eeqa}{\end{eqnarray}}
\newcommand{\pdg}{{\vphantom \dag}}
\newcommand{\dg}{{\dag}}
\newcommand{\bnabla}{{\boldsymbol \nabla}}

\begin{document}
\title{Chiral spin liquid from magnetic Wannier states}
\author{I. Panfilov}
\affiliation{Department of Physics and Astronomy, University of Waterloo, Waterloo, Ontario 
N2L 3G1, Canada}
\author{A. Patri}
\affiliation{Department of Physics and Astronomy, University of Waterloo, Waterloo, Ontario 
N2L 3G1, Canada}
\author{Kun Yang}
\affiliation{Department of Physics and National High Magnetic Field Laboratory, Florida State University, Tallahassee, Florida 32310, USA}
\author{A.A. Burkov}
\affiliation{Department of Physics and Astronomy, University of Waterloo, Waterloo, Ontario 
N2L 3G1, Canada} 
\affiliation{ITMO University, Saint Petersburg 197101, Russia}
\date{\today}
\begin{abstract}
We present a mapping of a two-dimensional system of interacting bosons in a strong perpendicular magnetic field to an 
equivalent system of interacting bosons on the square lattice in the absence of the field.  
The mapping utilizes a magnetic Bloch and the corresponding magnetic Wannier single-particle basis in the lowest Landau level. 
By construction, the ground states of the resulting model of interacting bosons on the square lattice are gapped fractionalized 
liquids or gapless Bose metal states with broken time reversal symmetry at specific rational filling fractions. 
\end{abstract}
\maketitle
\section{Introduction}
\label{sec:1}
Following the remarkable discovery of topological insulators (TI),~\cite{Hasan10,Qi11} electronic structure topology has been understood to be a
previously largely overlooked, but essential ingredient in our understanding of the phases of condensed matter.~\cite{Ryu15}
One of the reasons nontrivial electronic structure topology is of significant importance is that it is a purely quantum-mechanical phenomenon, with no 
classical analogs, yet, in many cases, has observable manifestations on macroscopic scales. 
This makes such phenomena not only interesting from the purely scientific viewpoint, but also potentially useful technologically. 

Quantum mechanical nature of the electrons in solids may also manifest on macroscopic scales through the electron-electron interactions,
well-known examples being the phenomena of magnetism and superconductivity. 
Perhaps particularly remarkable is the fractional quantum Hall effect (FQHE), where electrons effectively fractionalize and the low-energy quasiparticles 
are characterized by fractional quantum numbers and non-fermionic statistics. 
This amazing behavior is made possible by the interplay of the strong electron-electron interactions (kinetic energy being completely quenched by the 
magnetic field), and the nontrivial topology of the individual Landau levels. 

An important question is whether such phenomena are unique to the system of two-dimensional electrons in a strong perpendicular magnetic field, 
or they are more general and may be found in other systems where both interactions and nontrivial electronic structure topology are present. 
This question was first raised in the seminal paper of Kalmeyer and Laughlin,~\cite{Laughlin87} who pointed out strong similarities between the physics of FQHE and 
the resonating valence bond theory~\cite{Baskaran87} of spin-liquid states in Mott insulators.~\cite{Balents10}
The interest in this issue was reinvigorated recently, after the discovery of TI, which demonstrated that nontrivial electronic structure topology is quite common among heavy-element 
compounds with strong spin-orbit interactions.~\cite{Pesin10}
This gives one some hope that analogs of FQHE may be found in crystalline materials with nontrivial electronic structure topology in the absence of an external magnetic field (such a hypothetical material may be called a fractional Chern insulator). 

There has by now been a significant amount of work on fractional Chern insulators, see Refs.~\onlinecite{Sheng11-1,Chamon11,Bernevig11,Sun11,Wen11,Fiete11,Sondhi12,Bernevig12,Sheng12,Roy14,Daghofer12,Sheng11-2,Mudry11,Murthy12,Jain12,Lauchli12,Chamon12,XLQi11,Regnault12,Moller12,XLQi13,Thomale15} for an incomplete list. 
The purpose of this article is to derive, somewhat in the spirit of the Kalmeyer and Laughlin paper,~\cite{Laughlin87} a mapping between a model of interacting bosons in the lowest Landau level (LLL), and a lattice model of bosons in the absence of an external magnetic field (but with broken time reversal symmetry), with the lattice filling identical to the LLL filling factor. 
By construction, the ground states of this lattice model are equivalent to the ground states of interacting bosons in the LLL, i.e. may be fractionalized liquids or Bose metals~\cite{Motrunich07,Motrunich08} with broken time reversal symmetry at specific rational filling factors. 
While most of the calculations, presented below, may be carried out for a model of interacting electrons in the LLL just as well, we choose interacting bosons, having in mind potential realizations in magnetic systems,~\cite{Balents10} or cold atoms in optical lattices.~\cite{Bloch05} Some work along these lines, but valid only at high boson filling factors, has already been done by one of us.~\cite{Burkov10}
In this paper we present a more complete analysis, containing points related to the LLL topology, overlooked in 
Ref.~\onlinecite{Burkov10}, but crucially important at low boson filling factors, at which fractionalized liquid states are realized. 

The rest of the paper is organized as follows. 
In section~\ref{sec:2} we introduce our model of interacting bosons in two dimensions (2D) in the presence of a strong perpendicular magnetic field, such that the LLL projection may be used. 
We introduce a magnetic Bloch and magnetic Wannier single particle basis in the LLL, following the procedure, first described by Rashba {\em et al.}~\cite{Rashba97}
The advantage of this particular realization of the magnetic Wannier states is that they have the fastest possible decay rate for such states, $1/r^2$, in all directions, and have the full symmetry of the Bravais lattice.~\cite{Thouless84}
In section~\ref{sec:3} we derive a representation of the density operator in the magnetic Bloch basis and point out some of its most important properties. 
Focusing on long-wavelength density modes, we perform a gradient expansion of the density operator and derive a simplified expression, valid in the long-wavelength limit. 
In section~\ref{sec:4}, using the results obtained in the previous sections, we rewrite the Hamiltonian of interacting bosons in the LLL in the magnetic Wannier basis, using the 
long-wavelength expressions for the density operators. We show that in this long-wavelength limit the Hamiltonian has a very simple form, consisting only of a few distinct (long-range) terms. This provides a mapping between the Hamiltonian of interacting bosons in the LLL and a lattice boson Hamiltonian, with no explicitly present external magnetic field (broken time reversal symmetry is still explicit, however, since some of the terms in the Hamiltonian are complex).
We conclude in section~\ref{sec:5} with a discussion of our results and a brief summary. 

\section{Magnetic Bloch and magnetic Wannier bases in the LLL}
\label{sec:2}
We start from a model of bosons of charge $-e$, interacting via some two-body interaction potential in 2D 
(to be specified in more detail below), in the presence of a perpendicular magnetic field $\bB = B \hat z$. 
We will assume that the magnetic field is sufficiently strong, such that only states in the LLL are important. 

We will use the LLL basis of magnetic Wannier states, first introduced in Ref.~\onlinecite{Rashba97}. 
The advantage of this particular realization of the magnetic Wannier states is that they have the fastest 
decay rate at long distances, compatible with the nontrivial LLL topology, which is $1/r^2$.~\cite{Thouless84} They also are highly symmetric and allow, as will be demonstrated below, for 
the construction of Wannier Hamiltonians with the full symmetry of any 2D Bravais
lattice.

To construct this basis, we adopt the symmetric gauge ${\bf A} = \frac{1}{2} {\bf B} \times {\bf r}$, and
start from the zero-angular-momentum symmetric gauge orbital in the LLL
\beq
\label{eq:1}
c_0({\bf r}) = \frac{1}{\sqrt{2 \pi \ell^2}} e^{-\frac{r^2}{4 \ell^2}}, 
\eeq
where $\ell = \sqrt{c/e B}$ is the magnetic length and will use the $\hbar = 1$ units throughout. 
We then construct an overcomplete basis of the LLL orbitals by translating the $c_0({\bf r})$ orbital, 
localized at the origin, to sites of any 2D Bravais lattice with unit cell area $2 \pi \ell^2$. 
We will focus on the simplest case of the square lattice henceforth, as the latest geometry is unimportant 
here, see discussion of this point below. 
We obtain
\beqa
\label{eq:2}
c_{\bf m}({\bf r})&=&T_{m_x {\bf a}_x} T_{m_y {\bf a}_y} c_0({\bf r}) \nonumber \\
&=&\frac{(-1)^{m_x m_y}}{\sqrt{2 \pi \ell^2}} e^{-\frac{(\br - \br_{\bf m})^2}{4 \ell^2} + \frac{i}{2 \ell^2} 
\hat z \cdot (\br \times \br_{\bf m})}. 
\eeqa
Here 
\beq
\label{eq:3}
T_\bR = e^{-i \bR \cdot \left(\bp - \frac{e}{c} \bA \right)}, 
\eeq
is the magnetic translation operator in the symmetric gauge, $\ba_{x,y} = a \hat x, a\hat y$ 
are the primitive translation vectors of the square lattice with the lattice constant $a = \sqrt{2 \pi \ell^2}$, 
and $\bbm = (m_x, m_y)$ is a vector with integer components, labeling the lattice sites. 

The set of functions $c_\bbm(\br)$ is overcomplete by exactly one state, which is a consequence of 
the Perelomov identity~\cite{Perelomov}
\beq
\label{eq:4}
\sum_\bbm (-1)^{m_x + m_y} c_\bbm(\br) = 0.
\eeq
This property plays an important role in what follows. 

\begin{figure}[t]
  \includegraphics[width=7cm]{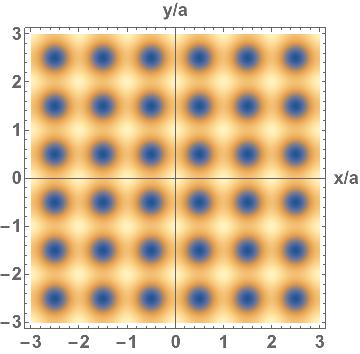}
  \caption{(Color online). Plot of $|\Psi_\bk(\br)|^2$ for $\bk = 0$. In general $|\Psi_\bk(\br)|^2$ has the form of a square Abrikosov vortex lattice, shifted with respect to the lattice, shown in the figure, by the vector $\ell^2 \hat z \times \bk$.}  
  \label{fig:abrikosov}
\end{figure}

The magnetic Bloch states may now be constructed as linear combinations of the LLL orbitals
$c_\bbm(\br)$ as
\beqa
\label{eq:5}
&&\Psi_\bk(\br) = \frac{1}{\sqrt{N \nu(\bk)}} \sum_\bbm c_\bbm(\br) e^{i \bk \cdot \br_\bbm} \nonumber \\
&=&\sqrt{\frac{2}{a^2 N \nu(\bk)}} e^{- \frac{\pi}{2 a^2} \br^2} e^{- \frac{a^2}{2 \pi} \left[k_y + \frac{\pi}{a^2} (x - i y) \right]^2}
\nonumber \\
&\times&\theta_3\left(\frac{k_+ a}{2}, e^{-\pi}\right) \theta_3\left(\frac{k_- a}{2} - \frac{i \pi}{a}(x - i y), e^{-\pi} \right). 
\eeqa
where $N = L_x L_y / 2 \pi \ell^2$ is the number of the magnetic flux quanta, contained in the sample 
area $L_x L_y$, $\theta_3(z,q)$ are Jacobi theta functions and $\nu(\bk)$ is needed to normalize the Bloch function to unity in the sample volume. 
Explicitly, the normalization factor is given by
\beqa
\label{eq:6}
\nu(\bk)&=&\sum_\bbm (-1)^{m_x m_y} e^{- \frac{\br_\bbm^2}{4 \ell^2}} e^{i  \bk \cdot \br_\bbm} \nonumber \\
&=&\sqrt{2} e^{- \frac{k_y^2 a^2}{2 \pi}} \theta_3\left(\frac{k_+ a}{2}, e^{-\pi} \right) 
\theta_3\left(\frac{k_- a}{2}, e^{-\pi}\right).
\eeqa
The probability density, corresponding to a magnetic Bloch state, $|\Psi_\bk(\br)|^2$, has the form of a square Abrikosov vortex lattice, as shown in Fig.~\ref{fig:abrikosov}. 

The function $\nu(\bk)$ is non-negative everywhere in the first Brillouin zone (BZ). 
As immediately follows from the Perelomov identity, Eq.~\eqref{eq:4}, $\nu(\bk)$ vanishes 
at the BZ corner $\bk_0 = (\pi/a, \pi/a)$, $\nu(\bk_0) = 0$. 
Near $\bk_0$, $\nu(\bk)$ behaves as
\beq
\label{eq:7}
\nu(\bk_0 + \bk) \approx \frac{\gamma}{2} \bk^2 a^2, 
\eeq
where 
\beq
\label{eq:8}
\gamma = - \frac{1}{2 a} \sum_\bbm (-1)^{m_x + m_y} c_\bbm(0) \br_\bbm^2, 
\eeq
is a positive constant of order unity. 
We will use the above results extensively later. 

The quantum geometry of the magnetic Bloch states, defined by Eq.~\eqref{eq:5}, turns out 
to be closely connected to the properties of the function $\nu(\bk)$. 
Defining a ``periodic part" of the Bloch function in the standard way as (it is periodic, but with respect 
to the magnetic translations, not ordinary ones) $u_\bk(\br) = e^{-i \bk \cdot \br} \Psi_\bk(\br)$,
and evaluating the Berry connection $\bA(\bk) = - i \langle u_\bk | \bnabla_\bk | u_\bk \rangle$, 
we obtain
\beq
\label{eq:9}
\bA(\bk) = \frac{1}{2} (\hat z \times \bnabla_\bk) \ln[\nu(\bk)].
\eeq
This may be particularly easily evaluated near the BZ corners. 
In this case, using Eq.~\eqref{eq:7}, we obtain
\beq
\label{eq:10}
\bA(\bk_0 + \bk) \approx \frac{\hat z \times \bk}{\bk^2}. 
\eeq
This expression is singular when $\bk \rightarrow 0$, which expresses the impossibility of choosing a smooth gauge for the Bloch functions in the LLL, due to the nonzero Chern number, to be calculated below. 
The BZ corner is where a ``Dirac string" must enter the first BZ, to ensure that the circulation of the 
Berry connection around the BZ boundary is equal to $2 \pi$. 
Evaluating the $z$-component of the Berry curvature, we obtain
\beq
\label{eq:11}
\Omega_z(\bk) = \bnabla_\bk \times \bA(\bk) = \frac{1}{2} \bnabla_\bk^2 \ln[\nu(\bk)]  = 
- \frac{a^2}{2 \pi},
\eeq
which follows immediately from Eq.~\eqref{eq:6}. 
The integral of $\Omega_z(\bk)$ over the BZ then gives the nontrivial Chern number of the LLL, 
as it should
\beq
\label{eq:12}
C = \frac{1}{2 \pi} \int_{-\pi/a}^{\pi/a} d k_x d k_y \Omega_z(\bk) = - 1. 
\eeq
This result may also be obtained using the expression Eq.~\eqref{eq:10} for the 
Berry connection near the BZ corners. 
If we evaluate the circulation of the Berry connection around the firs BZ boundary, it is clear that, 
due to the periodicity of the function $\nu(\bk)$ in the first BZ, only the singular points at the BZ 
corners will actually contribute to the circulation, see Fig.~\ref{fig:bz}. 
Using the Eq.~\eqref{eq:10}, one obtains
\beq
\label{eq:13}
\oint_{\partial BZ} \bA(\bk) \cdot d \bk = - 2 \pi, 
\eeq
which is equivalent to Eq.~\eqref{eq:12}. 

The magnetic Wannier states are related to the Bloch states in the standard way
\beq
\label{eq:14}
\Phi_\bbm(\br) = \frac{1}{\sqrt{N}} \sum_\bk \Psi_\bk(\br) e^{-i \bk \cdot \br_\bbm}. 
\eeq 
It is straightforward to show~\cite{Rashba97} that the divergence of the normalization factor of the Bloch wavefunction at the 
BZ corner 
\beq
\label{eq:14.1}
\frac{1}{\sqrt{\nu(\bk_0 + \bk)}} \sim \frac{1}{k}, 
\eeq
leads to power-law $1/r^2$ tail in the long-distance decay of the Wannier states $\Phi_\bbm(\br)$. 
Nonetheless, the functions $\Phi_\bbm(\br)$  form a complete orthonormal set of states, since the Bloch 
function $\Psi_{\bk_0}(\br)$ is still well-defined, the singularity, in the form of a momentum-space vortex, only existing in its phase
\beq
\label{eq:14.2}
\Psi_{\bk_0 + \bk}(\br) = \frac{i e^{i \phi_\bk}}{\sqrt{2 N \gamma}} \sum_\bbm (-1)^{m_x + m_y} (m_x - i m_y) c_\bbm (\br), 
\eeq
where $\bk \rightarrow 0$ and $\phi_\bk$ is the azimuthal angle of the vector $\bk$. This phase singularity is again a consequence of the Dirac string, as in Eq.~\eqref{eq:10}. 
Since the functions $\Phi_\bbm(\br)$ form a complete orthonormal set of states, the question of the LLL Hamiltonian in the magnetic Wannier basis is also well-defined. 

\section{Density operator in the magnetic Bloch and Wannier bases}
\label{sec:3}
In this section we will construct the density operator in the Bloch and Wannier bases, introduced 
in the previous section. 
As is well-known, the peculiar algebra 
(Girvin-MacDonald-Platzman, or GMP algebra)~\cite{GMP} of the LLL-projected density operator plays a crucial role in the appearance of the fractional quantum Hall liquid states in the LLL. 
It is thus important to understand how this algebra is realized when the density operator is written 
in the magnetic Bloch and Wannier bases. 

Evaluating the Fourier transform of the LLL-projected density operator, one obtains
\beqa
\label{eq:15}
\varrho(\bq)&=&\int d^2 r \Psi^\dg(\br) \Psi^\pdg(\br) e^{- i \bq \cdot \br} \nonumber \\
&=& e^{- \frac{q^2 a^2}{4 \pi}} \sum_\bk \frac{\nu\left(\bk + \frac{\bq}{2} - \frac{i}{2} \hat z \times \bq \right)}{\sqrt{\nu(\bk + \bq) 
\nu(\bk)}} b^\dg_{\bk} b^\pdg_{\bk + \bq}.
\eeqa
This expression may be simplified further either using Jacobi theta function identities or invoking properties of the 
Bloch functions $\Psi_\bk(\br)$. 
We will take the second route as it is more transparent.

The property of the Bloch functions we will use is that they are fully determined, up to a $\bk$-dependent phase factor, 
by their zeroes, which form a square Abrikosov vortex lattice with the lattice constant $a$, as shown in Fig.~\ref{fig:abrikosov}.  
Mathematically, this statement may be expressed in the form of the following relation~\cite{Burkov10}
\beq
\label{eq:16}
\Psi_{\bk + \bq}(\br) = e^{i \gamma(\bk+ \bq, \bk)} e^{\frac{i}{2} \bq \cdot \br} \Psi_\bk(\br - \ell^2 \hat z \times \bq), 
\eeq
where $e^{i \bq \cdot \br /2}$ is an Aharonov-Bohm phase factor and 
\beqa
\label{eq:17}
e^{i \gamma(\bk + \bq, \bk)}&=&\int d^2 r \Psi_{\bk + \bq}(\br) \Psi^*_\bk(\br - \ell^2 \hat z \times \bq) e^{-\frac{i}{2} \bq \cdot \br} \nonumber \\
&=&e^{-\frac{q^2 a^2}{8 \pi}} \frac{\nu\left(\bk + \frac{\bq}{2} - \frac{i}{2} \hat z \times \bq \right)}{\sqrt{\nu(\bk + \bq) \nu(\bk)}}.
\eeqa  
This immediately gives
\beq
\label{eq:18}
\gamma(\bk + \bq, \bk) = \textrm{Im} \ln \nu\left(\bk + \frac{\bq}{2} - \frac{i}{2} \hat z \times \bq\right), 
\eeq
or, equivalently
\beq
\label{eq:19}
\frac{\nu\left(\bk + \frac{\bq}{2} - \frac{i}{2} \hat z \times \bq \right)}{\sqrt{\nu(\bk + \bq) \nu(\bk)}} =
e^{\frac{q^2 a^2}{8\pi}} e^{i \textrm{Im} \ln \nu\left(\bk + \frac{\bq}{2} - \frac{i}{2} \hat z \times \bq\right)}. 
\eeq
The physical meaning of the phase $\gamma(\bk + \bq, \bk)$ is the momentum-space Berry phase, accumulated upon adiabatic evolution of the Bloch state from $\bk$ to $\bk + \bq$ (strictly speaking, a path in the first BZ needs to be specified for this identification to be precise, but this will not be necessary for our purposes). 

Thus we finally obtain the following expression for the density operator
\beq
\label{eq:20}
\varrho(\bq) = e^{- \frac{q^2 a^2}{8 \pi}} \sum_\bk e^{i \gamma(\bk + \bq, \bk)}
 b^\dg_{\bk} b^\pdg_{\bk + \bq} \equiv e^{- \frac{q^2 a^2}{8 \pi}} \bar \varrho(\bq). 
 \eeq
Using Eqs.~\eqref{eq:11} and \eqref{eq:16} it is straightforward to show that the density operators $\bar \varrho(\bq)$ satisfy the GMP algebra 
\beq
\label{eq:21}
[\bar \varrho(\bq), \bar \varrho(\bq')] = - 2 i \sin\left[\frac{a^2}{4 \pi} \hat z \cdot (\bq \times \bq') \right] \bar \varrho(\bq + \bq'),
\eeq
as they should.

 To make further progress we will assume that $q\, a$ may be taken to be small, i.e. only the long-wavelength density modes 
 are of interest to us. 
 This might, perhaps, be justified using renormalization-group-type arguments, although it is not easy in the present case, as we are interested in gapped fractionalized liquid phases with short correlation length. 
 We will thus take a more simple-minded approach here and assume the interparticle interaction potential has a long, but finite, range $\xi \gg a$. 
Contribution of the density modes with $q > 1/\xi$ is then suppressed naturally, without renormalization. 
This also gives us a natural small parameter $a/\xi$, which we will use to control our theory. 
Extensive earlier studies of the FQHE in finite-width quantum well systems~\cite{DasSarma90} indicate that $\xi$ may be safely 
taken to be as long as $10$ magnetic lengths. 

\begin{figure}[t]
  \includegraphics[width=7cm]{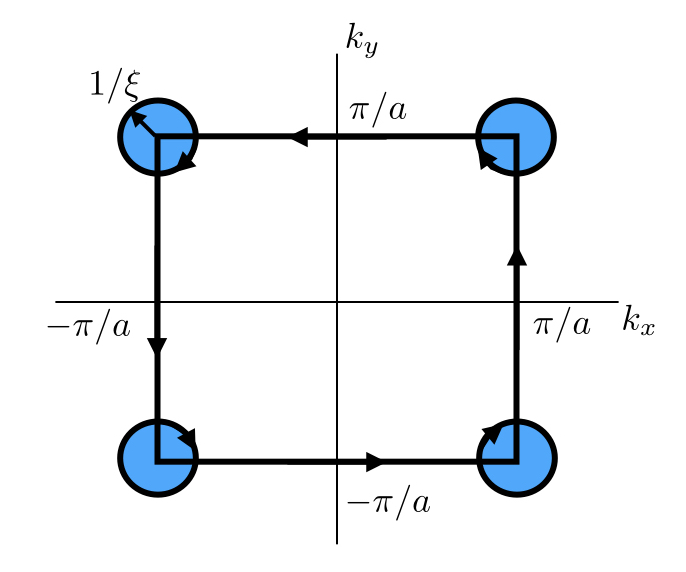}
  \caption{(Color online). First Brillouin zone with corner patches shown by shaded circles of radius $1/\xi$.
  Circulation of the Berry connection vector around the BZ boundary, excluding the corners, as shown by arrows, 
 gives the Chern number $C = -1$.}  
  \label{fig:bz}
\end{figure}

Taking $q \,a \ll 1$, the Berry phase given by Eq.~\eqref{eq:18}
is an analytic function of $\bq$ everywhere, except in a circular patch of radius $1/\xi \ll 1/a$ around the BZ corner, 
see Fig.~\ref{fig:bz}. 
This is again a consequence of the Dirac string, entering the BZ at the corner. 
Let us analyze the behavior of $\gamma(\bk + \bq, \bk)$ near the BZ corner 
 $\bk = \bk_0$ in detail. 
 Using Eq.~\eqref{eq:7}, one obtains
\beq
\label{eq:23}
\textrm{Im} \ln \nu\left(\bk_0 + \bk + \frac{\bq}{2} - \frac{i}{2} \hat z \times \bq\right) \approx \textrm{atan} \frac{(\hat z \times \bk) \cdot \bq} {\bk^2 + \bk \cdot \bq}.
\eeq
The meaning of Eq.~\eqref{eq:23} is simply the azimuthal angle between the directions of the vector $\bk + \bq$ and vector
 $\bk$. 
 This may be viewed as a direct consequence of Eq.~\eqref{eq:14.2}.  
Outside of the BZ corner patch, where $q < k$, this gives
\beqa
\label{eq:24}
&&\textrm{Im} \ln \nu\left(\bk_0 + \bk + \frac{\bq}{2} - \frac{i}{2} \hat z \times \bq\right) \nonumber \\
&\approx&\frac{\hat z \times \bk}{\bk^2} \cdot \bq 
\approx \bA(\bk_0 + \bk) \cdot \bq, 
\eeqa

The above discussion makes it clear that, in general, $\bar \varrho(\bq)$ is a nonanalytic function of $\bq$ in the vicinity 
of $\bq = 0$ and thus may not be expanded in Taylor series with respect to $\bq$.
However, as will be seen below, the nonanalyticity appears explicitly only when one goes beyond the first order in $\bq$, or, 
in other words, $\bnabla \bar \varrho(\bq \rightarrow 0)$ is finite, even though all the higher gradients are not.  
Thus, the gradient expansion of $\bar \varrho(\bq)$ does exist, if it is restricted to terms of up to first order in $q a$, or $a/\xi$. 
Expanding to only this order, we thus obtain
\beq
\label{eq:28}
\bar \varrho(\bq) \approx \sum_\bk e^{i \bA(\bk) \cdot \bq} b^\dg_\bk b^\pdg_{\bk + \bq} \approx \sum_\bk \left[1 + i \bA(\bk) \cdot \bq\right] b^\dg_{\bk} b^\pdg_{\bk + \bq}. 
\eeq
This is an expression for the LLL-projected density operator to leading nontrivial order in the small parameter $a/\xi$.

\section{LLL Hamiltonian in the magnetic Wannier basis}
\label{sec:4}
We now rewrite Eq.~\eqref{eq:28} in the magnetic Wannier basis using
\beq
\label{eq:28.5}
b^\dg_\bk = \frac{1}{\sqrt{N}}\sum_\bbm b^\dg_\bbm e^{i \bk \cdot \br_\bbm}. 
\eeq
One obtains
\beqa
\label{eq:29}
&&\bar \varrho(\bq) = \sum_\bbm e^{- i \bq \cdot \br_{\bbm}} b^\dg_\bbm b^\pdg_\bbm \nonumber \\
&+&\frac{i}{N} \sum_{\bbm \bbm'} 
\sum_\bk \bA(\bk) \cdot \bq e^{i \bk \cdot (\br_\bbm - \br_{\bbm'})} e^{- i \bq \cdot \br_{\bbm'}} b^\dg_\bbm b^\pdg_{\bbm'}. 
\eeqa
Even though the Berry connection $\bA(\bk)$ is singular in the limit $\bk \rightarrow \bk_0$ due to the presence of the 
Dirac string, the integral over $\bk$ in Eq.~\eqref{eq:29} still converges (but divergent terms will appear if expansion to higher 
orders in $\bq$ is attempted). 
However, due to the divergence of the Berry connection, the main contribution to the integral over $\bk$ at long distances, i.e. when $|\br_\bbm - \br_{\bbm'}| \gg a$, comes from the vicinity of the BZ corner. 
This also makes sense physically since, as discussed at the end of Section~\ref{sec:2}, it is the contribution of vicinity of the BZ corner that leads to the $1/r^2$ tail of the Wannier function $\Phi_\bbm(\br)$. 
In this case $\bA(\bk)$ may be approximated by Eq.~\eqref{eq:10} and the integral over $\bk$ in Eq.~\eqref{eq:29} is then 
easily evaluated analytically. 
We obtain
\beqa
\label{eq:30}
&&\bar \varrho(\bq) = \sum_\bbm e^{- i \bq \cdot \br_\bbm} b^\dg_\bbm b^\pdg_\bbm - \frac{a^2}{2 \pi} \sum_{\bbm \bbm'} 
\bq \cdot \frac{\hat z \times (\br_\bbm - \br_{\bbm'})}{(\br_\bbm - \br_{\bbm'})^2} \nonumber \\
&\times&e^{i \bk_0 \cdot (\br_\bbm - \br_{\bbm'})} e^{- i \bq \cdot \br_{\bbm'}}
b^\dg_\bbm b^\pdg_{\bbm'}.
\eeqa
The oscillating phase factor $e^{i \bk_0\cdot (\br _\bbm - \br_{\bbm'})}$ may be eliminated by a gauge transformation of the boson creation and annihilation operators 
\beq
\label{eq:31}
b_\bbm e^{-i \bk_0 \cdot \br_\bbm} \rightarrow b_\bbm, 
\eeq
and we will ignore this factor henceforth. 

The interacting-boson Hamiltonian, projected to the LLL, is given by
\beq
\label{eq:32}
H = \frac{1}{2 L_x L_y} \sum_\bq U(\bq) e^{-\frac{\bq^2 a^2}{4 \pi}} \bar \varrho(\bq) \bar \varrho(- \bq), 
\eeq
where $U(\bq)$ is the Fourier transform of the interparticle interaction potential. 
In accordance with the discussion above we take $U(\bq)$ to be negligible when $q > 1/\xi$ 
and equal to a constant $U(\bq) = U \xi^2$ when $q < 1/\xi$, where $U$ has dimensions of energy. 
In this case the integral over $\bq$ in Eq.~\eqref{eq:32} is easily done analytically. 
Restricting ourselves to only the terms of zeroth and first order in the small parameter $a/\xi$, we 
obtain $H = H_0 + H_1$, where 
\beq
\label{eq:33}
H_0 = U_0 \sum_{\bbm \bn} \frac{\textrm{J}_1(|\br_\bbm - \br_\bn|/\xi)}{|\br_\bbm - \br_\bn|/\xi}
b^\dg_\bbm b^\pdg_\bbm b^\dg_\bn b^\pdg_\bn, 
\eeq
and 
\beqa
\label{eq:34}
&&H_1 =  i \frac{a U_1}{\xi} \sum_{\bbm \bbm' \bn} \hat z \cdot \frac{ a \,\xi \, (\br_\bbm - \br_{\bbm'}) \times 
(\br_{\bbm'} - \br_\bn)}{|\br_{\bbm'} - \br_\bn|^2 |\br_\bbm - \br_{\bbm'}|^2} \nonumber \\
&\times&\textrm{J}_2(|\br_{\bbm'} - \br_\bn |/\xi) b^\dg_\bbm b^\pdg_{\bbm'} b^\dg_\bn b^\pdg_\bn, 
\eeqa
where $J_{1,2}$ are Bessel functions of the corresponding order, $U_0 = U/4 \pi$, and $U_1 = U/4 \pi^2$. 
Eqs.~\eqref{eq:33} and \eqref{eq:34} constitute the main result of our paper. 

It may be useful, especially for possible future numerical studies of this model, to extend it by introducing an ordinary 
kinetic energy term (hopping) for the bosons. 
Physically this may be achieved by adding an external potential, with exactly the same periodicity as the square lattice, 
formed by the magnetic Wannier state centers. 
This potential would introduce a boson hopping term of the form
\beq
\label{eq:35}
H_2 = - t \sum_{\langle {\bbm \bbm'} \rangle} e^{i \bk \cdot (\br_\bbm - \br_{\bbm'})} b^\dg_\bbm b^\pdg_{\bbm'}. 
\eeq
Here hopping is assumed to be restricted to the nearest-neighbor pairs of sites, $t > 0$ and the phase factor 
$e^{i \bk \cdot (\br_\bbm - \br_{\bbm'})}$ 
depends on the location of the Wannier orbital center within the unit cell of the physical square lattice. 
The ground states of $H_0$ correspond to the bosons condensing into one of the Bloch states $\Psi_\bk(\br)$  and
forming an Abrikosov vortex lattice state.
The value of $\bk$ determines the location of the vortex cores of the Abrikosov lattice relative to the Wannier orbital centers. 

The full Hamiltonian $H = H_0 + H_1 + H_2$ will then contain both an ordinary superfluid phase, when the $H_2$ term 
is dominant, and the fractionalized chiral liquid phases when $H_0 + H_1$ is dominant. 
This may be generalized even 
further by allowing the coupling constants $U_0$ and $U_1$ to be independent, which will also introduce ordinary Mott insulator 
phases with broken translational symmetry when $U_0 \gg t, U_1$.

\section{Discussion and conclusions}
\label{sec:5}
In the previous sections we have derived, using a controlled expansion in the small parameter $a/\xi$, 
a magnetic Wannier state representation of interacting bosons in the LLL. 
This Hamiltonian describes interacting bosons on a square lattice (the lattice geometry does not play a role 
here, as discussed below). The magnetic field does not enter explicitly 
in this Hamiltonian, but time reversal symmetry is still explicitly broken in the $H_1$ part of the Hamiltonian. 
The form of the $H_1$ term is a direct consequence of Eq.~\eqref{eq:28} and in this sense it may be regarded as a realization of  the GMP algebra, satisfied by the LLL-projected density operators. 

By construction, the ground states of the lattice Hamiltonian must be the same as the ground states of interacting bosons in the LLL, 
which include gapped fractionalized quantum Hall liquids at some rational filling fractions, e.g. at filling factor $1/2$. 
In this sense, the Hamiltonian given by Eqs.~\eqref{eq:33},\eqref{eq:34},\eqref{eq:35} may be regarded as a parent Hamiltonian of 
chiral spin liquids (although time reversal is broken explicitly here). 
Chiral spin liquids have attracted considerable attention,~\cite{Laughlin87,Wen89,Yang93,Moore05,Kivelson07,Thomale07} particularly due to recent work demonstrating they may be realized in spin-$1/2$ antiferromagnets on the kagome lattice.~\cite{Yang93,Trebst14,Sheng14,Sheng15-1,Sheng15-2,He15,Moessner15,Bieri15}
While in our model time reversal symmetry is broken explicitly, it may still be a useful starting point for constructing 
models in which it is broken spontaneously. 

The expression for the kinetic part of the Hamiltonian, Eq.~\eqref{eq:34}, reveals several features, which are presumably important to achieve a chiral spin liquid. 
First, $H_1$ has the form of a correlated hopping Hamiltonian, where the boson hopping amplitude from site $\bbm$ 
to $\bbm'$ depends on the boson density at site $\bn$. 
This reminds one, not accidentally, of flux attachment.~\cite{Pollmann15}
Second, both $H_0$ and $H_1$ terms are long-range, exhibiting power-law decay.  
This is a direct consequence of the long-range $1/r^2$ tail of the Wannier functions $\Phi_\bbm(\br)$, which in turn is 
rooted in the nontrivial topology of the LLL. 
This feature is also shared with previous constructions of parent Hamiltonians for chiral spin liquids,~\cite{Moore05,Thomale07}
employing mapping to the LLL in some form.
The power law decay in Eqs.~\eqref{eq:33}, \eqref{eq:34} is the fastest possible for this type of construction, 
since we are using the Wannier states with the fastest possible decay rate in all directions.  

In our construction of the magnetic Wannier basis we have chosen to place the Wannier orbitals on the sites of a square 
lattice. This choice is of course arbitrary: any 2D Bravais lattice would work just as well. 
This may appear strange since the lattice geometry naturally plays a crucial role within the prevailing 
paradigm in the search for quantum spin liquid physics, that of geometrically frustrated magnets. 
In our model, however, lattice geometry plays no role at all due to the long range nature of the interactions. 
The choice of the square lattice was thus dictated only by its simplicity.
Our construction, however, may be repeated on any 2D Bravais lattice with identical results. 

In conclusion, we have provided a derivation of a lattice Hamiltonian which, by construction, will have 
gapped fractionalized liquid ground states at certain rational filling fractional filling fractions, such as $1/2$. 
The construction employs magnetic Wannier states, which are highly symmetric (the Hamiltonian has the full 
symmetry of the Bravais lattice used) and have the fastest possible decay rate, compatible with the nontrivial topology of the 
LLL, i.e. $1/r^2$. It would be interesting to confirm our construction by an explicit numerical solution of 
Eqs.~\eqref{eq:33},\eqref{eq:34},\eqref{eq:35}. 

\begin{acknowledgments}
We acknowledge a useful discussion with D. Hawthorn. Financial support was provided by Natural Sciences and Engineering Research Council (NSERC) of Canada (IP, AP and AAB) and by the 
National Science Foundation through grants DMR-1157490 and DMR-1442366 (KY). 
\end{acknowledgments}
\bibliography{references}

\begin{thebibliography}{51}%
\makeatletter
\providecommand \@ifxundefined [1]{%
 \@ifx{#1\undefined}
}%
\providecommand \@ifnum [1]{%
 \ifnum #1\expandafter \@firstoftwo
 \else \expandafter \@secondoftwo
 \fi
}%
\providecommand \@ifx [1]{%
 \ifx #1\expandafter \@firstoftwo
 \else \expandafter \@secondoftwo
 \fi
}%
\providecommand \natexlab [1]{#1}%
\providecommand \enquote  [1]{``#1''}%
\providecommand \bibnamefont  [1]{#1}%
\providecommand \bibfnamefont [1]{#1}%
\providecommand \citenamefont [1]{#1}%
\providecommand \href@noop [0]{\@secondoftwo}%
\providecommand \href [0]{\begingroup \@sanitize@url \@href}%
\providecommand \@href[1]{\@@startlink{#1}\@@href}%
\providecommand \@@href[1]{\endgroup#1\@@endlink}%
\providecommand \@sanitize@url [0]{\catcode `\\12\catcode `\$12\catcode
  `\&12\catcode `\#12\catcode `\^12\catcode `\_12\catcode `\%12\relax}%
\providecommand \@@startlink[1]{}%
\providecommand \@@endlink[0]{}%
\providecommand \url  [0]{\begingroup\@sanitize@url \@url }%
\providecommand \@url [1]{\endgroup\@href {#1}{\urlprefix }}%
\providecommand \urlprefix  [0]{URL }%
\providecommand \Eprint [0]{\href }%
\providecommand \doibase [0]{http://dx.doi.org/}%
\providecommand \selectlanguage [0]{\@gobble}%
\providecommand \bibinfo  [0]{\@secondoftwo}%
\providecommand \bibfield  [0]{\@secondoftwo}%
\providecommand \translation [1]{[#1]}%
\providecommand \BibitemOpen [0]{}%
\providecommand \bibitemStop [0]{}%
\providecommand \bibitemNoStop [0]{.\EOS\space}%
\providecommand \EOS [0]{\spacefactor3000\relax}%
\providecommand \BibitemShut  [1]{\csname bibitem#1\endcsname}%
\let\auto@bib@innerbib\@empty
\bibitem [{\citenamefont {Hasan}\ and\ \citenamefont {Kane}(2010)}]{Hasan10}%
  \BibitemOpen
  \bibfield  {author} {\bibinfo {author} {\bibfnamefont {M.~Z.}\ \bibnamefont
  {Hasan}}\ and\ \bibinfo {author} {\bibfnamefont {C.~L.}\ \bibnamefont
  {Kane}},\ }\href {\doibase 10.1103/RevModPhys.82.3045} {\bibfield  {journal}
  {\bibinfo  {journal} {Rev. Mod. Phys.}\ }\textbf {\bibinfo {volume} {82}},\
  \bibinfo {pages} {3045} (\bibinfo {year} {2010})}\BibitemShut {NoStop}%
\bibitem [{\citenamefont {Qi}\ and\ \citenamefont {Zhang}(2011)}]{Qi11}%
  \BibitemOpen
  \bibfield  {author} {\bibinfo {author} {\bibfnamefont {X.-L.}\ \bibnamefont
  {Qi}}\ and\ \bibinfo {author} {\bibfnamefont {S.-C.}\ \bibnamefont {Zhang}},\
  }\href {\doibase 10.1103/RevModPhys.83.1057} {\bibfield  {journal} {\bibinfo
  {journal} {Rev. Mod. Phys.}\ }\textbf {\bibinfo {volume} {83}},\ \bibinfo
  {pages} {1057} (\bibinfo {year} {2011})}\BibitemShut {NoStop}%
\bibitem [{\citenamefont {{Chiu}}\ \emph {et~al.}(2015)\citenamefont {{Chiu}},
  \citenamefont {{Teo}}, \citenamefont {{Schnyder}},\ and\ \citenamefont
  {{Ryu}}}]{Ryu15}%
  \BibitemOpen
  \bibfield  {author} {\bibinfo {author} {\bibfnamefont {C.-K.}\ \bibnamefont
  {{Chiu}}}, \bibinfo {author} {\bibfnamefont {J.~C.~Y.}\ \bibnamefont
  {{Teo}}}, \bibinfo {author} {\bibfnamefont {A.~P.}\ \bibnamefont
  {{Schnyder}}}, \ and\ \bibinfo {author} {\bibfnamefont {S.}~\bibnamefont
  {{Ryu}}},\ }\href@noop {} {\bibfield  {journal} {\bibinfo  {journal} {ArXiv
  e-prints}\ } (\bibinfo {year} {2015})},\ \Eprint
  {http://arxiv.org/abs/1505.03535} {arXiv:1505.03535 [cond-mat.mes-hall]}
  \BibitemShut {NoStop}%
\bibitem [{\citenamefont {Kalmeyer}\ and\ \citenamefont
  {Laughlin}(1987)}]{Laughlin87}%
  \BibitemOpen
  \bibfield  {author} {\bibinfo {author} {\bibfnamefont {V.}~\bibnamefont
  {Kalmeyer}}\ and\ \bibinfo {author} {\bibfnamefont {R.~B.}\ \bibnamefont
  {Laughlin}},\ }\href {\doibase 10.1103/PhysRevLett.59.2095} {\bibfield
  {journal} {\bibinfo  {journal} {Phys. Rev. Lett.}\ }\textbf {\bibinfo
  {volume} {59}},\ \bibinfo {pages} {2095} (\bibinfo {year}
  {1987})}\BibitemShut {NoStop}%
\bibitem [{\citenamefont {Baskaran}\ \emph {et~al.}(1987)\citenamefont
  {Baskaran}, \citenamefont {Zou},\ and\ \citenamefont
  {Anderson}}]{Baskaran87}%
  \BibitemOpen
  \bibfield  {author} {\bibinfo {author} {\bibfnamefont {G.}~\bibnamefont
  {Baskaran}}, \bibinfo {author} {\bibfnamefont {Z.}~\bibnamefont {Zou}}, \
  and\ \bibinfo {author} {\bibfnamefont {P.}~\bibnamefont {Anderson}},\ }\href
  {\doibase http://dx.doi.org/10.1016/0038-1098(87)90642-9} {\bibfield
  {journal} {\bibinfo  {journal} {Solid State Communications}\ }\textbf
  {\bibinfo {volume} {63}},\ \bibinfo {pages} {973 } (\bibinfo {year}
  {1987})}\BibitemShut {NoStop}%
\bibitem [{\citenamefont {Balents}(2010)}]{Balents10}%
  \BibitemOpen
  \bibfield  {author} {\bibinfo {author} {\bibfnamefont {L.}~\bibnamefont
  {Balents}},\ }\href@noop {} {\bibfield  {journal} {\bibinfo  {journal}
  {Nature}\ }\textbf {\bibinfo {volume} {464}},\ \bibinfo {pages} {199}
  (\bibinfo {year} {2010})}\BibitemShut {NoStop}%
\bibitem [{\citenamefont {Pesin}\ and\ \citenamefont
  {Balents}(2010)}]{Pesin10}%
  \BibitemOpen
  \bibfield  {author} {\bibinfo {author} {\bibfnamefont {D.}~\bibnamefont
  {Pesin}}\ and\ \bibinfo {author} {\bibfnamefont {L.}~\bibnamefont
  {Balents}},\ }\href {http://dx.doi.org/10.1038/nphys1606} {\bibfield
  {journal} {\bibinfo  {journal} {Nat Phys}\ }\textbf {\bibinfo {volume} {6}},\
  \bibinfo {pages} {376} (\bibinfo {year} {2010})}\BibitemShut {NoStop}%
\bibitem [{\citenamefont {Sheng}\ \emph {et~al.}(2011)\citenamefont {Sheng},
  \citenamefont {Gu}, \citenamefont {Sun},\ and\ \citenamefont
  {Sheng}}]{Sheng11-1}%
  \BibitemOpen
  \bibfield  {author} {\bibinfo {author} {\bibfnamefont {D.~N.}\ \bibnamefont
  {Sheng}}, \bibinfo {author} {\bibfnamefont {Z.-C.}\ \bibnamefont {Gu}},
  \bibinfo {author} {\bibfnamefont {K.}~\bibnamefont {Sun}}, \ and\ \bibinfo
  {author} {\bibfnamefont {L.}~\bibnamefont {Sheng}},\ }\href
  {http://dx.doi.org/10.1038/ncomms1380} {\bibfield  {journal} {\bibinfo
  {journal} {Nat Commun}\ }\textbf {\bibinfo {volume} {2}},\ \bibinfo {pages}
  {389} (\bibinfo {year} {2011})}\BibitemShut {NoStop}%
\bibitem [{\citenamefont {Neupert}\ \emph
  {et~al.}(2011{\natexlab{a}})\citenamefont {Neupert}, \citenamefont {Santos},
  \citenamefont {Chamon},\ and\ \citenamefont {Mudry}}]{Chamon11}%
  \BibitemOpen
  \bibfield  {author} {\bibinfo {author} {\bibfnamefont {T.}~\bibnamefont
  {Neupert}}, \bibinfo {author} {\bibfnamefont {L.}~\bibnamefont {Santos}},
  \bibinfo {author} {\bibfnamefont {C.}~\bibnamefont {Chamon}}, \ and\ \bibinfo
  {author} {\bibfnamefont {C.}~\bibnamefont {Mudry}},\ }\href {\doibase
  10.1103/PhysRevLett.106.236804} {\bibfield  {journal} {\bibinfo  {journal}
  {Phys. Rev. Lett.}\ }\textbf {\bibinfo {volume} {106}},\ \bibinfo {pages}
  {236804} (\bibinfo {year} {2011}{\natexlab{a}})}\BibitemShut {NoStop}%
\bibitem [{\citenamefont {Regnault}\ and\ \citenamefont
  {Bernevig}(2011)}]{Bernevig11}%
  \BibitemOpen
  \bibfield  {author} {\bibinfo {author} {\bibfnamefont {N.}~\bibnamefont
  {Regnault}}\ and\ \bibinfo {author} {\bibfnamefont {B.~A.}\ \bibnamefont
  {Bernevig}},\ }\href {\doibase 10.1103/PhysRevX.1.021014} {\bibfield
  {journal} {\bibinfo  {journal} {Phys. Rev. X}\ }\textbf {\bibinfo {volume}
  {1}},\ \bibinfo {pages} {021014} (\bibinfo {year} {2011})}\BibitemShut
  {NoStop}%
\bibitem [{\citenamefont {Sun}\ \emph {et~al.}(2011)\citenamefont {Sun},
  \citenamefont {Gu}, \citenamefont {Katsura},\ and\ \citenamefont
  {Das~Sarma}}]{Sun11}%
  \BibitemOpen
  \bibfield  {author} {\bibinfo {author} {\bibfnamefont {K.}~\bibnamefont
  {Sun}}, \bibinfo {author} {\bibfnamefont {Z.}~\bibnamefont {Gu}}, \bibinfo
  {author} {\bibfnamefont {H.}~\bibnamefont {Katsura}}, \ and\ \bibinfo
  {author} {\bibfnamefont {S.}~\bibnamefont {Das~Sarma}},\ }\href {\doibase
  10.1103/PhysRevLett.106.236803} {\bibfield  {journal} {\bibinfo  {journal}
  {Phys. Rev. Lett.}\ }\textbf {\bibinfo {volume} {106}},\ \bibinfo {pages}
  {236803} (\bibinfo {year} {2011})}\BibitemShut {NoStop}%
\bibitem [{\citenamefont {Tang}\ \emph {et~al.}(2011)\citenamefont {Tang},
  \citenamefont {Mei},\ and\ \citenamefont {Wen}}]{Wen11}%
  \BibitemOpen
  \bibfield  {author} {\bibinfo {author} {\bibfnamefont {E.}~\bibnamefont
  {Tang}}, \bibinfo {author} {\bibfnamefont {J.-W.}\ \bibnamefont {Mei}}, \
  and\ \bibinfo {author} {\bibfnamefont {X.-G.}\ \bibnamefont {Wen}},\ }\href
  {\doibase 10.1103/PhysRevLett.106.236802} {\bibfield  {journal} {\bibinfo
  {journal} {Phys. Rev. Lett.}\ }\textbf {\bibinfo {volume} {106}},\ \bibinfo
  {pages} {236802} (\bibinfo {year} {2011})}\BibitemShut {NoStop}%
\bibitem [{\citenamefont {Hu}\ \emph {et~al.}(2011)\citenamefont {Hu},
  \citenamefont {Kargarian},\ and\ \citenamefont {Fiete}}]{Fiete11}%
  \BibitemOpen
  \bibfield  {author} {\bibinfo {author} {\bibfnamefont {X.}~\bibnamefont
  {Hu}}, \bibinfo {author} {\bibfnamefont {M.}~\bibnamefont {Kargarian}}, \
  and\ \bibinfo {author} {\bibfnamefont {G.~A.}\ \bibnamefont {Fiete}},\ }\href
  {\doibase 10.1103/PhysRevB.84.155116} {\bibfield  {journal} {\bibinfo
  {journal} {Phys. Rev. B}\ }\textbf {\bibinfo {volume} {84}},\ \bibinfo
  {pages} {155116} (\bibinfo {year} {2011})}\BibitemShut {NoStop}%
\bibitem [{\citenamefont {Parameswaran}\ \emph {et~al.}(2012)\citenamefont
  {Parameswaran}, \citenamefont {Roy},\ and\ \citenamefont
  {Sondhi}}]{Sondhi12}%
  \BibitemOpen
  \bibfield  {author} {\bibinfo {author} {\bibfnamefont {S.~A.}\ \bibnamefont
  {Parameswaran}}, \bibinfo {author} {\bibfnamefont {R.}~\bibnamefont {Roy}}, \
  and\ \bibinfo {author} {\bibfnamefont {S.~L.}\ \bibnamefont {Sondhi}},\
  }\href {\doibase 10.1103/PhysRevB.85.241308} {\bibfield  {journal} {\bibinfo
  {journal} {Phys. Rev. B}\ }\textbf {\bibinfo {volume} {85}},\ \bibinfo
  {pages} {241308} (\bibinfo {year} {2012})}\BibitemShut {NoStop}%
\bibitem [{\citenamefont {Wu}\ \emph {et~al.}(2012{\natexlab{a}})\citenamefont
  {Wu}, \citenamefont {Bernevig},\ and\ \citenamefont {Regnault}}]{Bernevig12}%
  \BibitemOpen
  \bibfield  {author} {\bibinfo {author} {\bibfnamefont {Y.-L.}\ \bibnamefont
  {Wu}}, \bibinfo {author} {\bibfnamefont {B.~A.}\ \bibnamefont {Bernevig}}, \
  and\ \bibinfo {author} {\bibfnamefont {N.}~\bibnamefont {Regnault}},\ }\href
  {\doibase 10.1103/PhysRevB.85.075116} {\bibfield  {journal} {\bibinfo
  {journal} {Phys. Rev. B}\ }\textbf {\bibinfo {volume} {85}},\ \bibinfo
  {pages} {075116} (\bibinfo {year} {2012}{\natexlab{a}})}\BibitemShut
  {NoStop}%
\bibitem [{\citenamefont {Wang}\ \emph {et~al.}(2012)\citenamefont {Wang},
  \citenamefont {Yao}, \citenamefont {Gu}, \citenamefont {Gong},\ and\
  \citenamefont {Sheng}}]{Sheng12}%
  \BibitemOpen
  \bibfield  {author} {\bibinfo {author} {\bibfnamefont {Y.-F.}\ \bibnamefont
  {Wang}}, \bibinfo {author} {\bibfnamefont {H.}~\bibnamefont {Yao}}, \bibinfo
  {author} {\bibfnamefont {Z.-C.}\ \bibnamefont {Gu}}, \bibinfo {author}
  {\bibfnamefont {C.-D.}\ \bibnamefont {Gong}}, \ and\ \bibinfo {author}
  {\bibfnamefont {D.~N.}\ \bibnamefont {Sheng}},\ }\href {\doibase
  10.1103/PhysRevLett.108.126805} {\bibfield  {journal} {\bibinfo  {journal}
  {Phys. Rev. Lett.}\ }\textbf {\bibinfo {volume} {108}},\ \bibinfo {pages}
  {126805} (\bibinfo {year} {2012})}\BibitemShut {NoStop}%
\bibitem [{\citenamefont {Roy}(2014)}]{Roy14}%
  \BibitemOpen
  \bibfield  {author} {\bibinfo {author} {\bibfnamefont {R.}~\bibnamefont
  {Roy}},\ }\href {\doibase 10.1103/PhysRevB.90.165139} {\bibfield  {journal}
  {\bibinfo  {journal} {Phys. Rev. B}\ }\textbf {\bibinfo {volume} {90}},\
  \bibinfo {pages} {165139} (\bibinfo {year} {2014})}\BibitemShut {NoStop}%
\bibitem [{\citenamefont {Venderbos}\ \emph {et~al.}(2012)\citenamefont
  {Venderbos}, \citenamefont {Kourtis}, \citenamefont {van~den Brink},\ and\
  \citenamefont {Daghofer}}]{Daghofer12}%
  \BibitemOpen
  \bibfield  {author} {\bibinfo {author} {\bibfnamefont {J.~W.~F.}\
  \bibnamefont {Venderbos}}, \bibinfo {author} {\bibfnamefont {S.}~\bibnamefont
  {Kourtis}}, \bibinfo {author} {\bibfnamefont {J.}~\bibnamefont {van~den
  Brink}}, \ and\ \bibinfo {author} {\bibfnamefont {M.}~\bibnamefont
  {Daghofer}},\ }\href {\doibase 10.1103/PhysRevLett.108.126405} {\bibfield
  {journal} {\bibinfo  {journal} {Phys. Rev. Lett.}\ }\textbf {\bibinfo
  {volume} {108}},\ \bibinfo {pages} {126405} (\bibinfo {year}
  {2012})}\BibitemShut {NoStop}%
\bibitem [{\citenamefont {Wang}\ \emph {et~al.}(2011)\citenamefont {Wang},
  \citenamefont {Gu}, \citenamefont {Gong},\ and\ \citenamefont
  {Sheng}}]{Sheng11-2}%
  \BibitemOpen
  \bibfield  {author} {\bibinfo {author} {\bibfnamefont {Y.-F.}\ \bibnamefont
  {Wang}}, \bibinfo {author} {\bibfnamefont {Z.-C.}\ \bibnamefont {Gu}},
  \bibinfo {author} {\bibfnamefont {C.-D.}\ \bibnamefont {Gong}}, \ and\
  \bibinfo {author} {\bibfnamefont {D.~N.}\ \bibnamefont {Sheng}},\ }\href
  {\doibase 10.1103/PhysRevLett.107.146803} {\bibfield  {journal} {\bibinfo
  {journal} {Phys. Rev. Lett.}\ }\textbf {\bibinfo {volume} {107}},\ \bibinfo
  {pages} {146803} (\bibinfo {year} {2011})}\BibitemShut {NoStop}%
\bibitem [{\citenamefont {Neupert}\ \emph
  {et~al.}(2011{\natexlab{b}})\citenamefont {Neupert}, \citenamefont {Santos},
  \citenamefont {Ryu}, \citenamefont {Chamon},\ and\ \citenamefont
  {Mudry}}]{Mudry11}%
  \BibitemOpen
  \bibfield  {author} {\bibinfo {author} {\bibfnamefont {T.}~\bibnamefont
  {Neupert}}, \bibinfo {author} {\bibfnamefont {L.}~\bibnamefont {Santos}},
  \bibinfo {author} {\bibfnamefont {S.}~\bibnamefont {Ryu}}, \bibinfo {author}
  {\bibfnamefont {C.}~\bibnamefont {Chamon}}, \ and\ \bibinfo {author}
  {\bibfnamefont {C.}~\bibnamefont {Mudry}},\ }\href {\doibase
  10.1103/PhysRevB.84.165107} {\bibfield  {journal} {\bibinfo  {journal} {Phys.
  Rev. B}\ }\textbf {\bibinfo {volume} {84}},\ \bibinfo {pages} {165107}
  (\bibinfo {year} {2011}{\natexlab{b}})}\BibitemShut {NoStop}%
\bibitem [{\citenamefont {Murthy}\ and\ \citenamefont
  {Shankar}(2012)}]{Murthy12}%
  \BibitemOpen
  \bibfield  {author} {\bibinfo {author} {\bibfnamefont {G.}~\bibnamefont
  {Murthy}}\ and\ \bibinfo {author} {\bibfnamefont {R.}~\bibnamefont
  {Shankar}},\ }\href {\doibase 10.1103/PhysRevB.86.195146} {\bibfield
  {journal} {\bibinfo  {journal} {Phys. Rev. B}\ }\textbf {\bibinfo {volume}
  {86}},\ \bibinfo {pages} {195146} (\bibinfo {year} {2012})}\BibitemShut
  {NoStop}%
\bibitem [{\citenamefont {Wu}\ \emph {et~al.}(2012{\natexlab{b}})\citenamefont
  {Wu}, \citenamefont {Jain},\ and\ \citenamefont {Sun}}]{Jain12}%
  \BibitemOpen
  \bibfield  {author} {\bibinfo {author} {\bibfnamefont {Y.-H.}\ \bibnamefont
  {Wu}}, \bibinfo {author} {\bibfnamefont {J.~K.}\ \bibnamefont {Jain}}, \ and\
  \bibinfo {author} {\bibfnamefont {K.}~\bibnamefont {Sun}},\ }\href {\doibase
  10.1103/PhysRevB.86.165129} {\bibfield  {journal} {\bibinfo  {journal} {Phys.
  Rev. B}\ }\textbf {\bibinfo {volume} {86}},\ \bibinfo {pages} {165129}
  (\bibinfo {year} {2012}{\natexlab{b}})}\BibitemShut {NoStop}%
\bibitem [{\citenamefont {Liu}\ \emph {et~al.}(2012)\citenamefont {Liu},
  \citenamefont {Bergholtz}, \citenamefont {Fan},\ and\ \citenamefont
  {L\"auchli}}]{Lauchli12}%
  \BibitemOpen
  \bibfield  {author} {\bibinfo {author} {\bibfnamefont {Z.}~\bibnamefont
  {Liu}}, \bibinfo {author} {\bibfnamefont {E.~J.}\ \bibnamefont {Bergholtz}},
  \bibinfo {author} {\bibfnamefont {H.}~\bibnamefont {Fan}}, \ and\ \bibinfo
  {author} {\bibfnamefont {A.~M.}\ \bibnamefont {L\"auchli}},\ }\href {\doibase
  10.1103/PhysRevLett.109.186805} {\bibfield  {journal} {\bibinfo  {journal}
  {Phys. Rev. Lett.}\ }\textbf {\bibinfo {volume} {109}},\ \bibinfo {pages}
  {186805} (\bibinfo {year} {2012})}\BibitemShut {NoStop}%
\bibitem [{\citenamefont {Grushin}\ \emph {et~al.}(2012)\citenamefont
  {Grushin}, \citenamefont {Neupert}, \citenamefont {Chamon},\ and\
  \citenamefont {Mudry}}]{Chamon12}%
  \BibitemOpen
  \bibfield  {author} {\bibinfo {author} {\bibfnamefont {A.~G.}\ \bibnamefont
  {Grushin}}, \bibinfo {author} {\bibfnamefont {T.}~\bibnamefont {Neupert}},
  \bibinfo {author} {\bibfnamefont {C.}~\bibnamefont {Chamon}}, \ and\ \bibinfo
  {author} {\bibfnamefont {C.}~\bibnamefont {Mudry}},\ }\href {\doibase
  10.1103/PhysRevB.86.205125} {\bibfield  {journal} {\bibinfo  {journal} {Phys.
  Rev. B}\ }\textbf {\bibinfo {volume} {86}},\ \bibinfo {pages} {205125}
  (\bibinfo {year} {2012})}\BibitemShut {NoStop}%
\bibitem [{\citenamefont {Qi}(2011)}]{XLQi11}%
  \BibitemOpen
  \bibfield  {author} {\bibinfo {author} {\bibfnamefont {X.-L.}\ \bibnamefont
  {Qi}},\ }\href {\doibase 10.1103/PhysRevLett.107.126803} {\bibfield
  {journal} {\bibinfo  {journal} {Phys. Rev. Lett.}\ }\textbf {\bibinfo
  {volume} {107}},\ \bibinfo {pages} {126803} (\bibinfo {year}
  {2011})}\BibitemShut {NoStop}%
\bibitem [{\citenamefont {Wu}\ \emph {et~al.}(2012{\natexlab{c}})\citenamefont
  {Wu}, \citenamefont {Regnault},\ and\ \citenamefont {Bernevig}}]{Regnault12}%
  \BibitemOpen
  \bibfield  {author} {\bibinfo {author} {\bibfnamefont {Y.-L.}\ \bibnamefont
  {Wu}}, \bibinfo {author} {\bibfnamefont {N.}~\bibnamefont {Regnault}}, \ and\
  \bibinfo {author} {\bibfnamefont {B.~A.}\ \bibnamefont {Bernevig}},\ }\href
  {\doibase 10.1103/PhysRevB.86.085129} {\bibfield  {journal} {\bibinfo
  {journal} {Phys. Rev. B}\ }\textbf {\bibinfo {volume} {86}},\ \bibinfo
  {pages} {085129} (\bibinfo {year} {2012}{\natexlab{c}})}\BibitemShut
  {NoStop}%
\bibitem [{\citenamefont {Scaffidi}\ and\ \citenamefont
  {M\"oller}(2012)}]{Moller12}%
  \BibitemOpen
  \bibfield  {author} {\bibinfo {author} {\bibfnamefont {T.}~\bibnamefont
  {Scaffidi}}\ and\ \bibinfo {author} {\bibfnamefont {G.}~\bibnamefont
  {M\"oller}},\ }\href {\doibase 10.1103/PhysRevLett.109.246805} {\bibfield
  {journal} {\bibinfo  {journal} {Phys. Rev. Lett.}\ }\textbf {\bibinfo
  {volume} {109}},\ \bibinfo {pages} {246805} (\bibinfo {year}
  {2012})}\BibitemShut {NoStop}%
\bibitem [{\citenamefont {Jian}\ and\ \citenamefont {Qi}(2013)}]{XLQi13}%
  \BibitemOpen
  \bibfield  {author} {\bibinfo {author} {\bibfnamefont {C.-M.}\ \bibnamefont
  {Jian}}\ and\ \bibinfo {author} {\bibfnamefont {X.-L.}\ \bibnamefont {Qi}},\
  }\href {\doibase 10.1103/PhysRevB.88.165134} {\bibfield  {journal} {\bibinfo
  {journal} {Phys. Rev. B}\ }\textbf {\bibinfo {volume} {88}},\ \bibinfo
  {pages} {165134} (\bibinfo {year} {2013})}\BibitemShut {NoStop}%
\bibitem [{\citenamefont {Claassen}\ \emph {et~al.}(2015)\citenamefont
  {Claassen}, \citenamefont {Lee}, \citenamefont {Thomale}, \citenamefont
  {Qi},\ and\ \citenamefont {Devereaux}}]{Thomale15}%
  \BibitemOpen
  \bibfield  {author} {\bibinfo {author} {\bibfnamefont {M.}~\bibnamefont
  {Claassen}}, \bibinfo {author} {\bibfnamefont {C.~H.}\ \bibnamefont {Lee}},
  \bibinfo {author} {\bibfnamefont {R.}~\bibnamefont {Thomale}}, \bibinfo
  {author} {\bibfnamefont {X.-L.}\ \bibnamefont {Qi}}, \ and\ \bibinfo {author}
  {\bibfnamefont {T.~P.}\ \bibnamefont {Devereaux}},\ }\href {\doibase
  10.1103/PhysRevLett.114.236802} {\bibfield  {journal} {\bibinfo  {journal}
  {Phys. Rev. Lett.}\ }\textbf {\bibinfo {volume} {114}},\ \bibinfo {pages}
  {236802} (\bibinfo {year} {2015})}\BibitemShut {NoStop}%
\bibitem [{\citenamefont {Motrunich}\ and\ \citenamefont
  {Fisher}(2007)}]{Motrunich07}%
  \BibitemOpen
  \bibfield  {author} {\bibinfo {author} {\bibfnamefont {O.~I.}\ \bibnamefont
  {Motrunich}}\ and\ \bibinfo {author} {\bibfnamefont {M.~P.~A.}\ \bibnamefont
  {Fisher}},\ }\href {\doibase 10.1103/PhysRevB.75.235116} {\bibfield
  {journal} {\bibinfo  {journal} {Phys. Rev. B}\ }\textbf {\bibinfo {volume}
  {75}},\ \bibinfo {pages} {235116} (\bibinfo {year} {2007})}\BibitemShut
  {NoStop}%
\bibitem [{\citenamefont {Sheng}\ \emph {et~al.}(2008)\citenamefont {Sheng},
  \citenamefont {Motrunich}, \citenamefont {Trebst}, \citenamefont {Gull},\
  and\ \citenamefont {Fisher}}]{Motrunich08}%
  \BibitemOpen
  \bibfield  {author} {\bibinfo {author} {\bibfnamefont {D.~N.}\ \bibnamefont
  {Sheng}}, \bibinfo {author} {\bibfnamefont {O.~I.}\ \bibnamefont
  {Motrunich}}, \bibinfo {author} {\bibfnamefont {S.}~\bibnamefont {Trebst}},
  \bibinfo {author} {\bibfnamefont {E.}~\bibnamefont {Gull}}, \ and\ \bibinfo
  {author} {\bibfnamefont {M.~P.~A.}\ \bibnamefont {Fisher}},\ }\href {\doibase
  10.1103/PhysRevB.78.054520} {\bibfield  {journal} {\bibinfo  {journal} {Phys.
  Rev. B}\ }\textbf {\bibinfo {volume} {78}},\ \bibinfo {pages} {054520}
  (\bibinfo {year} {2008})}\BibitemShut {NoStop}%
\bibitem [{\citenamefont {Bloch}(2005)}]{Bloch05}%
  \BibitemOpen
  \bibfield  {author} {\bibinfo {author} {\bibfnamefont {I.}~\bibnamefont
  {Bloch}},\ }\href {http://dx.doi.org/10.1038/nphys138} {\bibfield  {journal}
  {\bibinfo  {journal} {Nat Phys}\ }\textbf {\bibinfo {volume} {1}},\ \bibinfo
  {pages} {23} (\bibinfo {year} {2005})}\BibitemShut {NoStop}%
\bibitem [{\citenamefont {Burkov}(2010)}]{Burkov10}%
  \BibitemOpen
  \bibfield  {author} {\bibinfo {author} {\bibfnamefont {A.~A.}\ \bibnamefont
  {Burkov}},\ }\href {\doibase 10.1103/PhysRevB.81.125111} {\bibfield
  {journal} {\bibinfo  {journal} {Phys. Rev. B}\ }\textbf {\bibinfo {volume}
  {81}},\ \bibinfo {pages} {125111} (\bibinfo {year} {2010})}\BibitemShut
  {NoStop}%
\bibitem [{\citenamefont {Rashba}\ \emph {et~al.}(1997)\citenamefont {Rashba},
  \citenamefont {Zhukov},\ and\ \citenamefont {Efros}}]{Rashba97}%
  \BibitemOpen
  \bibfield  {author} {\bibinfo {author} {\bibfnamefont {E.~I.}\ \bibnamefont
  {Rashba}}, \bibinfo {author} {\bibfnamefont {L.~E.}\ \bibnamefont {Zhukov}},
  \ and\ \bibinfo {author} {\bibfnamefont {A.~L.}\ \bibnamefont {Efros}},\
  }\href {\doibase 10.1103/PhysRevB.55.5306} {\bibfield  {journal} {\bibinfo
  {journal} {Phys. Rev. B}\ }\textbf {\bibinfo {volume} {55}},\ \bibinfo
  {pages} {5306} (\bibinfo {year} {1997})}\BibitemShut {NoStop}%
\bibitem [{\citenamefont {Thouless}(1984)}]{Thouless84}%
  \BibitemOpen
  \bibfield  {author} {\bibinfo {author} {\bibfnamefont {D.~J.}\ \bibnamefont
  {Thouless}},\ }\href@noop {} {\bibfield  {journal} {\bibinfo  {journal} {J.
  Phys. C: Solid State Phys.}\ }\textbf {\bibinfo {volume} {17}},\ \bibinfo
  {pages} {L325} (\bibinfo {year} {1984})}\BibitemShut {NoStop}%
\bibitem [{\citenamefont {Perelomov}(1971)}]{Perelomov}%
  \BibitemOpen
  \bibfield  {author} {\bibinfo {author} {\bibfnamefont {A.~M.}\ \bibnamefont
  {Perelomov}},\ }\href@noop {} {\bibfield  {journal} {\bibinfo  {journal}
  {Teor. Mat. Fiz.}\ }\textbf {\bibinfo {volume} {6}},\ \bibinfo {pages} {213}
  (\bibinfo {year} {1971})}\BibitemShut {NoStop}%
\bibitem [{\citenamefont {Girvin}\ \emph {et~al.}(1986)\citenamefont {Girvin},
  \citenamefont {MacDonald},\ and\ \citenamefont {Platzman}}]{GMP}%
  \BibitemOpen
  \bibfield  {author} {\bibinfo {author} {\bibfnamefont {S.~M.}\ \bibnamefont
  {Girvin}}, \bibinfo {author} {\bibfnamefont {A.~H.}\ \bibnamefont
  {MacDonald}}, \ and\ \bibinfo {author} {\bibfnamefont {P.~M.}\ \bibnamefont
  {Platzman}},\ }\href {\doibase 10.1103/PhysRevB.33.2481} {\bibfield
  {journal} {\bibinfo  {journal} {Phys. Rev. B}\ }\textbf {\bibinfo {volume}
  {33}},\ \bibinfo {pages} {2481} (\bibinfo {year} {1986})}\BibitemShut
  {NoStop}%
\bibitem [{\citenamefont {He}\ \emph {et~al.}(1990)\citenamefont {He},
  \citenamefont {Zhang}, \citenamefont {Xie},\ and\ \citenamefont
  {Das~Sarma}}]{DasSarma90}%
  \BibitemOpen
  \bibfield  {author} {\bibinfo {author} {\bibfnamefont {S.}~\bibnamefont
  {He}}, \bibinfo {author} {\bibfnamefont {F.~C.}\ \bibnamefont {Zhang}},
  \bibinfo {author} {\bibfnamefont {X.~C.}\ \bibnamefont {Xie}}, \ and\
  \bibinfo {author} {\bibfnamefont {S.}~\bibnamefont {Das~Sarma}},\ }\href
  {\doibase 10.1103/PhysRevB.42.11376} {\bibfield  {journal} {\bibinfo
  {journal} {Phys. Rev. B}\ }\textbf {\bibinfo {volume} {42}},\ \bibinfo
  {pages} {11376} (\bibinfo {year} {1990})}\BibitemShut {NoStop}%
\bibitem [{\citenamefont {Wen}\ \emph {et~al.}(1989)\citenamefont {Wen},
  \citenamefont {Wilczek},\ and\ \citenamefont {Zee}}]{Wen89}%
  \BibitemOpen
  \bibfield  {author} {\bibinfo {author} {\bibfnamefont {X.~G.}\ \bibnamefont
  {Wen}}, \bibinfo {author} {\bibfnamefont {F.}~\bibnamefont {Wilczek}}, \ and\
  \bibinfo {author} {\bibfnamefont {A.}~\bibnamefont {Zee}},\ }\href {\doibase
  10.1103/PhysRevB.39.11413} {\bibfield  {journal} {\bibinfo  {journal} {Phys.
  Rev. B}\ }\textbf {\bibinfo {volume} {39}},\ \bibinfo {pages} {11413}
  (\bibinfo {year} {1989})}\BibitemShut {NoStop}%
\bibitem [{\citenamefont {Yang}\ \emph {et~al.}(1993)\citenamefont {Yang},
  \citenamefont {Warman},\ and\ \citenamefont {Girvin}}]{Yang93}%
  \BibitemOpen
  \bibfield  {author} {\bibinfo {author} {\bibfnamefont {K.}~\bibnamefont
  {Yang}}, \bibinfo {author} {\bibfnamefont {L.~K.}\ \bibnamefont {Warman}}, \
  and\ \bibinfo {author} {\bibfnamefont {S.~M.}\ \bibnamefont {Girvin}},\
  }\href {\doibase 10.1103/PhysRevLett.70.2641} {\bibfield  {journal} {\bibinfo
   {journal} {Phys. Rev. Lett.}\ }\textbf {\bibinfo {volume} {70}},\ \bibinfo
  {pages} {2641} (\bibinfo {year} {1993})}\BibitemShut {NoStop}%
\bibitem [{\citenamefont {Seidel}\ \emph {et~al.}(2005)\citenamefont {Seidel},
  \citenamefont {Fu}, \citenamefont {Lee}, \citenamefont {Leinaas},\ and\
  \citenamefont {Moore}}]{Moore05}%
  \BibitemOpen
  \bibfield  {author} {\bibinfo {author} {\bibfnamefont {A.}~\bibnamefont
  {Seidel}}, \bibinfo {author} {\bibfnamefont {H.}~\bibnamefont {Fu}}, \bibinfo
  {author} {\bibfnamefont {D.-H.}\ \bibnamefont {Lee}}, \bibinfo {author}
  {\bibfnamefont {J.~M.}\ \bibnamefont {Leinaas}}, \ and\ \bibinfo {author}
  {\bibfnamefont {J.}~\bibnamefont {Moore}},\ }\href {\doibase
  10.1103/PhysRevLett.95.266405} {\bibfield  {journal} {\bibinfo  {journal}
  {Phys. Rev. Lett.}\ }\textbf {\bibinfo {volume} {95}},\ \bibinfo {pages}
  {266405} (\bibinfo {year} {2005})}\BibitemShut {NoStop}%
\bibitem [{\citenamefont {Yao}\ and\ \citenamefont
  {Kivelson}(2007)}]{Kivelson07}%
  \BibitemOpen
  \bibfield  {author} {\bibinfo {author} {\bibfnamefont {H.}~\bibnamefont
  {Yao}}\ and\ \bibinfo {author} {\bibfnamefont {S.~A.}\ \bibnamefont
  {Kivelson}},\ }\href {\doibase 10.1103/PhysRevLett.99.247203} {\bibfield
  {journal} {\bibinfo  {journal} {Phys. Rev. Lett.}\ }\textbf {\bibinfo
  {volume} {99}},\ \bibinfo {pages} {247203} (\bibinfo {year}
  {2007})}\BibitemShut {NoStop}%
\bibitem [{\citenamefont {Schroeter}\ \emph {et~al.}(2007)\citenamefont
  {Schroeter}, \citenamefont {Kapit}, \citenamefont {Thomale},\ and\
  \citenamefont {Greiter}}]{Thomale07}%
  \BibitemOpen
  \bibfield  {author} {\bibinfo {author} {\bibfnamefont {D.~F.}\ \bibnamefont
  {Schroeter}}, \bibinfo {author} {\bibfnamefont {E.}~\bibnamefont {Kapit}},
  \bibinfo {author} {\bibfnamefont {R.}~\bibnamefont {Thomale}}, \ and\
  \bibinfo {author} {\bibfnamefont {M.}~\bibnamefont {Greiter}},\ }\href
  {\doibase 10.1103/PhysRevLett.99.097202} {\bibfield  {journal} {\bibinfo
  {journal} {Phys. Rev. Lett.}\ }\textbf {\bibinfo {volume} {99}},\ \bibinfo
  {pages} {097202} (\bibinfo {year} {2007})}\BibitemShut {NoStop}%
\bibitem [{\citenamefont {Bauer}\ \emph {et~al.}(2014)\citenamefont {Bauer},
  \citenamefont {Cincio}, \citenamefont {Keller}, \citenamefont {Dolfi},
  \citenamefont {Vidal}, \citenamefont {Trebst},\ and\ \citenamefont
  {Ludwig}}]{Trebst14}%
  \BibitemOpen
  \bibfield  {author} {\bibinfo {author} {\bibfnamefont {B.}~\bibnamefont
  {Bauer}}, \bibinfo {author} {\bibfnamefont {L.}~\bibnamefont {Cincio}},
  \bibinfo {author} {\bibfnamefont {B.~P.}\ \bibnamefont {Keller}}, \bibinfo
  {author} {\bibfnamefont {M.}~\bibnamefont {Dolfi}}, \bibinfo {author}
  {\bibfnamefont {G.}~\bibnamefont {Vidal}}, \bibinfo {author} {\bibfnamefont
  {S.}~\bibnamefont {Trebst}}, \ and\ \bibinfo {author} {\bibfnamefont
  {A.~W.~W.}\ \bibnamefont {Ludwig}},\ }\href
  {http://dx.doi.org/10.1038/ncomms6137} {\bibfield  {journal} {\bibinfo
  {journal} {Nat Commun}\ }\textbf {\bibinfo {volume} {5}},\ \bibinfo {pages}
  {5137} (\bibinfo {year} {2014})}\BibitemShut {NoStop}%
\bibitem [{\citenamefont {He}\ \emph {et~al.}(2014)\citenamefont {He},
  \citenamefont {Sheng},\ and\ \citenamefont {Chen}}]{Sheng14}%
  \BibitemOpen
  \bibfield  {author} {\bibinfo {author} {\bibfnamefont {Y.-C.}\ \bibnamefont
  {He}}, \bibinfo {author} {\bibfnamefont {D.~N.}\ \bibnamefont {Sheng}}, \
  and\ \bibinfo {author} {\bibfnamefont {Y.}~\bibnamefont {Chen}},\ }\href
  {\doibase 10.1103/PhysRevLett.112.137202} {\bibfield  {journal} {\bibinfo
  {journal} {Phys. Rev. Lett.}\ }\textbf {\bibinfo {volume} {112}},\ \bibinfo
  {pages} {137202} (\bibinfo {year} {2014})}\BibitemShut {NoStop}%
\bibitem [{\citenamefont {Zhu}\ \emph {et~al.}(2015)\citenamefont {Zhu},
  \citenamefont {Gong},\ and\ \citenamefont {Sheng}}]{Sheng15-1}%
  \BibitemOpen
  \bibfield  {author} {\bibinfo {author} {\bibfnamefont {W.}~\bibnamefont
  {Zhu}}, \bibinfo {author} {\bibfnamefont {S.~S.}\ \bibnamefont {Gong}}, \
  and\ \bibinfo {author} {\bibfnamefont {D.~N.}\ \bibnamefont {Sheng}},\ }\href
  {\doibase 10.1103/PhysRevB.92.014424} {\bibfield  {journal} {\bibinfo
  {journal} {Phys. Rev. B}\ }\textbf {\bibinfo {volume} {92}},\ \bibinfo
  {pages} {014424} (\bibinfo {year} {2015})}\BibitemShut {NoStop}%
\bibitem [{\citenamefont {Gong}\ \emph {et~al.}(2015)\citenamefont {Gong},
  \citenamefont {Zhu}, \citenamefont {Balents},\ and\ \citenamefont
  {Sheng}}]{Sheng15-2}%
  \BibitemOpen
  \bibfield  {author} {\bibinfo {author} {\bibfnamefont {S.-S.}\ \bibnamefont
  {Gong}}, \bibinfo {author} {\bibfnamefont {W.}~\bibnamefont {Zhu}}, \bibinfo
  {author} {\bibfnamefont {L.}~\bibnamefont {Balents}}, \ and\ \bibinfo
  {author} {\bibfnamefont {D.~N.}\ \bibnamefont {Sheng}},\ }\href {\doibase
  10.1103/PhysRevB.91.075112} {\bibfield  {journal} {\bibinfo  {journal} {Phys.
  Rev. B}\ }\textbf {\bibinfo {volume} {91}},\ \bibinfo {pages} {075112}
  (\bibinfo {year} {2015})}\BibitemShut {NoStop}%
\bibitem [{\citenamefont {He}\ and\ \citenamefont {Chen}(2015)}]{He15}%
  \BibitemOpen
  \bibfield  {author} {\bibinfo {author} {\bibfnamefont {Y.-C.}\ \bibnamefont
  {He}}\ and\ \bibinfo {author} {\bibfnamefont {Y.}~\bibnamefont {Chen}},\
  }\href {\doibase 10.1103/PhysRevLett.114.037201} {\bibfield  {journal}
  {\bibinfo  {journal} {Phys. Rev. Lett.}\ }\textbf {\bibinfo {volume} {114}},\
  \bibinfo {pages} {037201} (\bibinfo {year} {2015})}\BibitemShut {NoStop}%
\bibitem [{\citenamefont {He}\ \emph {et~al.}(2015{\natexlab{a}})\citenamefont
  {He}, \citenamefont {Bhattacharjee}, \citenamefont {Pollmann},\ and\
  \citenamefont {Moessner}}]{Moessner15}%
  \BibitemOpen
  \bibfield  {author} {\bibinfo {author} {\bibfnamefont {Y.-C.}\ \bibnamefont
  {He}}, \bibinfo {author} {\bibfnamefont {S.}~\bibnamefont {Bhattacharjee}},
  \bibinfo {author} {\bibfnamefont {F.}~\bibnamefont {Pollmann}}, \ and\
  \bibinfo {author} {\bibfnamefont {R.}~\bibnamefont {Moessner}},\ }\href
  {\doibase 10.1103/PhysRevLett.115.267209} {\bibfield  {journal} {\bibinfo
  {journal} {Phys. Rev. Lett.}\ }\textbf {\bibinfo {volume} {115}},\ \bibinfo
  {pages} {267209} (\bibinfo {year} {2015}{\natexlab{a}})}\BibitemShut
  {NoStop}%
\bibitem [{\citenamefont {Bieri}\ \emph {et~al.}(2015)\citenamefont {Bieri},
  \citenamefont {Messio}, \citenamefont {Bernu},\ and\ \citenamefont
  {Lhuillier}}]{Bieri15}%
  \BibitemOpen
  \bibfield  {author} {\bibinfo {author} {\bibfnamefont {S.}~\bibnamefont
  {Bieri}}, \bibinfo {author} {\bibfnamefont {L.}~\bibnamefont {Messio}},
  \bibinfo {author} {\bibfnamefont {B.}~\bibnamefont {Bernu}}, \ and\ \bibinfo
  {author} {\bibfnamefont {C.}~\bibnamefont {Lhuillier}},\ }\href {\doibase
  10.1103/PhysRevB.92.060407} {\bibfield  {journal} {\bibinfo  {journal} {Phys.
  Rev. B}\ }\textbf {\bibinfo {volume} {92}},\ \bibinfo {pages} {060407}
  (\bibinfo {year} {2015})}\BibitemShut {NoStop}%
\bibitem [{\citenamefont {He}\ \emph {et~al.}(2015{\natexlab{b}})\citenamefont
  {He}, \citenamefont {Bhattacharjee}, \citenamefont {Moessner},\ and\
  \citenamefont {Pollmann}}]{Pollmann15}%
  \BibitemOpen
  \bibfield  {author} {\bibinfo {author} {\bibfnamefont {Y.-C.}\ \bibnamefont
  {He}}, \bibinfo {author} {\bibfnamefont {S.}~\bibnamefont {Bhattacharjee}},
  \bibinfo {author} {\bibfnamefont {R.}~\bibnamefont {Moessner}}, \ and\
  \bibinfo {author} {\bibfnamefont {F.}~\bibnamefont {Pollmann}},\ }\href
  {\doibase 10.1103/PhysRevLett.115.116803} {\bibfield  {journal} {\bibinfo
  {journal} {Phys. Rev. Lett.}\ }\textbf {\bibinfo {volume} {115}},\ \bibinfo
  {pages} {116803} (\bibinfo {year} {2015}{\natexlab{b}})}\BibitemShut
  {NoStop}%
\end{thebibliography}%

\end{document}